**Title Page**

**Full Title**

Automatic detection of acute ischemic stroke using non-contrast computed tomography and two-stage deep learning model

**Authors and Affiliations**

Mizuho Nishio, M.D., Ph.D.[1], Sho Koyasu, M.D., Ph.D.[2,3], Shunjiro Noguchi, M.D.[4], Takao Kiguchi, M.D., Ph.D.[3], Kanako Nakatsu, M.D.[4], Thai Akasaka, M.D., Ph.D.[4], Hiroki Yamada, M.D., Ph.D.[3], Kyo Itoh, M.D., Ph.D.[4]

1 Department of Radiology, Kobe University Hospital, 7-5-2 Kusunoki-cho, Chuo-ku, Kobe 650-0017, Japan

2 Research Center for Advanced Science and Technology, The University of Tokyo, 4-6-1 Komaba, Meguro-ku, Tokyo, 153-8904, Japan.

3 Department of Radiology, Ichinomiyanishi Nishi Hospital, Ichinomiya, 1 Kaimei-hira, Ichinomiya, 494-0001 Japan.

4 Department of Radiology, Osaka Red Cross Hospital, 5-30 Fudegasakicho, Tennoji-ku, Osaka, 543-8555 Japan

*Corresponding author: Mizuho Nishio, M.D., Ph.D.

Department of Radiology, Kobe University Hospital

Tel: +81-78-382-6104. Fax: +81-78-382-6129.

e-mail: nishiomizuho@gmail.com

**Abstract**

**Background and Objective**: Currently, it is challenging to detect acute ischemic stroke-related (AIS) changes on computed tomography (CT) images. Therefore, we aimed to develop and evaluate an automatic AIS detection system involving a two-stage deep learning model.

**Methods**: We included 238 cases from two different institutions. AIS-related findings were annotated on each of the 238 sets of head CT images by referring to head magnetic resonance imaging (MRI) images in which an MRI examination was performed within 24 h following the CT scan. These 238 annotated cases were divided into a training set including 189 cases and test set including 49 cases. Subsequently, a two-stage deep learning detection model was constructed from the training set using the You Only Look Once v3 model and Visual Geometry Group 16 classification model. Then, the two-stage model performed the AIS detection process in the test set. To assess the detection model's results, a board-certified radiologist also evaluated the test set head CT images with and without the aid of the detection model. The sensitivity of AIS detection and number of false positives were calculated for the evaluation of the test set detection results. The sensitivity of the radiologist with and without the software detection results was compared using the McNemar test. A p-value of less than 0.05 was considered statistically significant.

**Results:** For the two-stage model and radiologist without and with the use of the software results, the sensitivity was 37.3%, 33.3%, and 41.3%, respectively, and the number of false positives per one case was 1.265, 0.327, and 0.388, respectively. On using the two-stage detection model's results, the board-certified radiologist's detection sensitivity significantly improved (p-value = 0.0313).

**Conclusions:** Our detection system involving the two-stage deep learning model significantly improved the radiologist's sensitivity in AIS detection.

**Introduction**

According to the World Health Organization, in 2016, stroke was found to be the second leading cause of death worldwide [1]. Generally, there are two types of strokes. The first type is ischemic stroke (brain infarction) that involves the blood vessels being blocked by thrombus or other causes. The other type is hemorrhagic stroke that, in most cases, is caused by the rupturing of blood vessels. Because computed tomography (CT) is a diagnostic modality that is characterized by its wide availability, low costs, and short acquisition time, it is often used as the first-line diagnostic technique for the evaluation of these types of stroke. Recently, frequently-employed treatment modalities for ischemic stroke have been endovascular thrombectomy [2] and the administration of an intravenous tissue plasminogen activator [3]. Ideally, for these treatments to be the most effective, a reduction is required in the duration of time from the onset of the disease to the commencement of treatment administration. Consequently, non-contrast CT is frequently used for the diagnosis of stroke.

Generally, for the diagnosis of acute ischemic stroke (AIS), magnetic resonance imaging (MRI), especially diffusion-weighted imaging (DWI), has a higher detectability level than CT [4,5]. However, MRI is inferior to CT in terms of its availability, costs, and acquisition time. Therefore, in clinical practice, in order to reduce the time duration from the disease onset until the start of treatment, non-contrast CT is normally used as the first choice diagnostic option for examining suspected stroke patients. Considering the current clinical practice situation and prevalence of CT scanners, an improvement is necessary in AIS detectability on CT images to improve the AIS-related clinical situation, especially in the case of emergency situations.

It is important to detect early ischemic changes resulting from AIS on non-contrast CT images as early as possible because a decision on whether or not a tissue plasminogen activator should be administered should be reached within 4.5 h following the disease onset [6]. However, the CT findings associated with early ischemic changes are subtle, and it is difficult to accurately identify them on CT images. For example, by using magnetic resonance imaging (MRI) as the gold standard for ischemic stroke detection, Akasaka et al. demonstrated that the mean sensitivity in stroke detection on non-contrast CT images among 14 radiologists was 26.5% [7]. Therefore, an automated system that could assist medical doctors in AIS detection on CT images might be valuable in a clinical situation.

Recently, the use of deep learning has been adopted in the field of medical image analysis and has yielded promising results [8–19]. For example, deep learning could be used to grade diabetic retinopathy [8] and to classify skin lesions as benign or malignant with an accuracy level equivalent to that of an expert [13]. In addition, deep learning has been used to discriminate benign/malignant lung nodules [17] and to detect pneumonia on chest x-ray images [12]. Concerning stroke, several studies have demonstrated that deep learning could successfully be used to detect intracranial hemorrhage and/or hemorrhagic stroke on CT images [10,11,16]. However, to the best of our knowledge, studies involving the detection of AIS using non-contrast CT and deep learning are sparse.

The purpose of the current study was to develop a deep learning software system for AIS detection on non-contrast CT images which was useful for radiologists. We believe that this software system could contribute to improving the clinical outcomes of AIS. For this purpose, we used a two-stage deep learning model system comprised of one deep learning model to detect AIS and another one to reduce false positives (FPs). The combination of the two models could lead to the reliable detection of AIS. As far as we know, an AIS detection system for non-contrast CT images has rarely been evaluated using an MRI-validated CT dataset. Because the detectability of CT is lower than that of MRI, a software system for AIS detection should be evaluated using an MRI-validated CT dataset. Therefore, in the current study, we also aimed to construct an MRI-validated CT dataset for AIS in order to develop and assess our detection system.

**Methods**

The institutional review boards of both Osaka Red Cross Hospital, Osaka, Japan and Ichinomiyanishi Nishi Hospital, Ichinomiya, Japan approved this retrospective study and waived the need for informed consent. The outline of our proposed method is presented in Figure 1. In the training phase, a training set from the MRI-validated CT dataset of AIS-related images was used for the construction of two deep learning models: one model for the detection of AIS and the other one for false positive reduction (FPR). In the deployment phase, a test set was used for assessing the combination of the results generated by the two models.

**Figure 1**

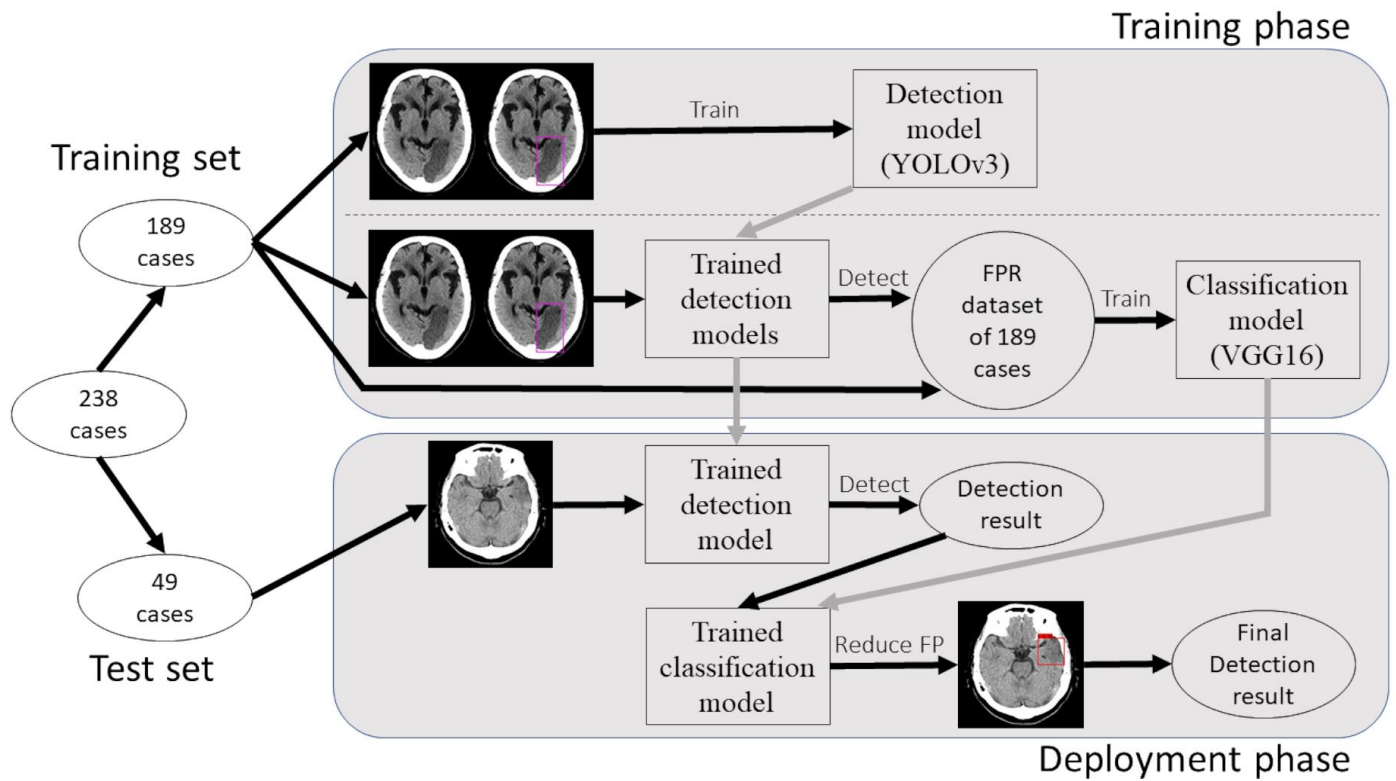

Outline of the proposed method for the detection of acute ischemic stroke

Abbreviations: false positive reduction (FPR), false positive (FP), You Only Look Once v3 (YOLOv3), Visual Geometry Group (VGG).

Dataset

From January 2017 to May 2018, we retrospectively obtained the data of a total of 238 consecutive patients (158 and 80 patients from Osaka Red Cross Hospital and Ichinomiyanishi Nishi Hospital, respectively) who were suspected of having AIS and who underwent head CT scans, followed by head MRI scans, including the capture of diffusion-weighted images (DWI), performed within 24 hours. Before the experimental procedures, patients were excluded if the slice angle used for their CT scans was not the same as that used for their MRI examination or if strong artifacts were observed on either the CT or MRI images. In total, 6114 head CT images were obtained from the 238 patients. The following are the imaging parameters that were used. In Osaka Red Cross Hospital, the CT (Aquilion One GENESIS Edition, Canon Medical Systems; Revolution GSI or Discovery 750HD, General Electric) slice thickness was 4 mm with a field of view (FOV) of 230 mm and 512×512 pixels. The slice thickness of the DWI (Intera 1.5T Nova Dual, Ingenia 3.0T HP, or Achieva 1.5T Nova Dual, Philips) was 6 mm with an FOV of 220 mm and 256×256 pixels. In Ichinomiyanishi Nishi Hospital, the CT (Aquilion ONE, Canon Medical Systems) slice thickness was 4 mm with an FOV of 220 mm and 512×512 pixels. The slice thickness of the DWI (MAGNETOM Verio, SIEMENS; EXCELART Vantage, Canon Medical Systems) was 6 mm with an FOV of 220 mm and 256×256 pixels. The dataset was annotated by a single board-certified radiologist. Using our in-house software, a rectangular bounding box was drawn on the head CT images for AIS detection. During the annotation process, the MRI images that corresponded to the head CT images were referred to side-by-side. Among the 6114 head CT images of the 238 cases, 1129 images depicting AIS were annotated (in several cases, one AIS was depicted on several CT images). The 238 annotated cases were divided into a training set consisting of 189 cases and a test set consisting of 49 cases.

Preprocessing

Before the construction of the deep learning model, the head CT images in the dataset were preprocessed. First, the CT values of the images were converted based on the brain window setup (window center: 30 Hounsfield unit (HU); window width: 60 HU). Subsequently, pixel value normalization was performed before the images were input into the model (all the pixel values were divided by 255). The images were subjected to online image augmentation techniques, such as horizontal flipping, cropping, and random rotation, while training the models.

Detection model for AIS

Two deep learning models were used to develop our two-stage deep learning detection system. The first model that was used for detecting AIS was based on the You Only Look Once v3 (YOLOv3) model [20]. The details associated with the YOLOv3 model are presented in the supplementary material. YOLOv3 is a one-stage detection algorithm that directly outputs the class probability and spatial coordinates as detection results. In our model, YOLOv3 outputted the probability and location of AIS. The Keras (https://keras.io/) with Tensorflow (https://www.tensorflow.org/) backends were used for the implementation of the YOLOv3 model. To improve the model's performance, a YOLOv3 model that was pre-trained with the Microsoft Common Objects in Context (MS-COCO) dataset was utilized (https://pjreddie.com/media/files/yolov3.weights), and transfer learning was performed in order to detect AIS. The network structure of YOLOv3 was not changed apart from its input and output.

    While training the YOLOv3 model, we divided the CT images from the training set into 169 training cases and 20 validation cases. The splitting of the training and validation cases was randomly performed 10 times, and training loss and validation loss were calculated for each split using the same hyperparameters [17]. The best model for evaluating the test set was selected based on the losses among the 10 models. The training procedure for the YOLOv3 model was divided into 3 steps. The optimizer, learning rate, and number of epochs were different for each step: Nadam, 0.001, and 40 for the first step; RMSprop, 0.0001, and 80 for the second step, and RMSprop, 0.00001, 280 for the third step, respectively. Early stopping and stepping learning rate decay were enabled only in the third

step. The patience of early stopping was 30. The patience and decay factor of the stepping learning rate decay were 10 and 0.1, respectively. The batch size for training the detection model was 8.

Using the 10 trained YOLOv3 models, an FPR dataset was constructed for use in the second deep-learning model from the training set comprising of 189 cases. The detection results obtained from the 10 models were classified into true positives (TPs) or FPs based on the annotation data of the training set. If the probability of AIS outputted by YOLOv3 was more than 0.02, the detection results were used for the FPR dataset; otherwise, the detection results were not used. If the Intersection over Union (IoU) between the detection result of the FPR dataset and annotation data was more than 0.3, the detection result was defined as a TP; otherwise, it was defined as an FP. In addition, the data (rectangular bounding box) of the training set in which AIS findings were annotated were added to the FPR dataset as TPs. IoU is defined in the supplementary material.

Classification model for false positive reduction

The second model was the classification model that was used for reducing FPs. The FPR model outputted the probability of whether the detection result obtained from the first model was a TP or an FP. The FPR dataset described in the previous subsection was used for training and evaluating the FPR model. A four-fold cross validation technique was used for the training and evaluation processes. The FPR model was derived from the Visual Geometry Group 16 (VGG16) convolutional neural network [21], and VGG16 was modified to perform transfer learning [17]. For the VGG16-based model, the input images should consist of three channels. In the current study, two FPR models were constructed: a one-slice CT image of the detection result was used as the input for one of the FPR models (the one-slice image was tripled for the 3-channel input), and three consecutive CT images of the detection result were used for the other FPR model. A schematic illustration of these two types of FPR models is presented in Figure 2. Because the detection result was represented as the rectangular bounding box, the image patch corresponding to the detection result was scaled as the input size of the FPR model (input size = 128×128×3). The optimizer, learning rate, and number of epochs of the FPR model were Nadam, 0.00002, and 20, respectively. The batch size for training the classification model was 100.

**Figure 2**

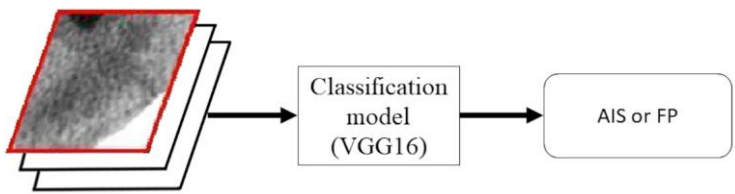

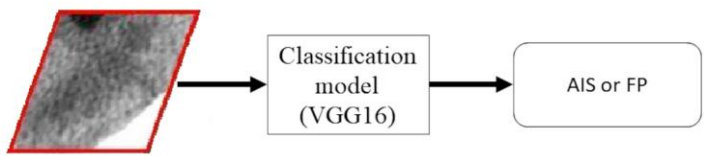

Schematic illustration of the two types of FPR models in the deployment phase

Abbreviations: false positive reduction (FPR), false positive (FP), acute ischemic stroke (AIS), Visual Geometry Group (VGG).

Evaluation

To assess the usefulness of our proposed model, we evaluated the results generated by the following three combinations of models: (I) the one-stage model (only YOLOv3), (II) two-stage model with the FPR model involving a one-slice image, and (III) two-stage model with the FPR model involving three-slice images. Based on the results of our preliminary experiments, an ensemble comprising of the four FPR models obtained from the four-fold cross validation was more effective than using the single FPR model. Thus, thereafter, we used the ensemble of four FPR models with the same hyperparameters.

The detection results of the one-stage model and two types of two-stage models were evaluated for the test set consisting of 49 cases, in which 75 AIS were annotated. The numbers of TPs and FPs of the models were counted for the 49 cases with 75 AIS, and the sensitivity (recall), FPs per case (FPC), precision, and F1 score were calculated. For the three combinations of models, the threshold for probability was selected based on the F1 score of the three combinations. Recall, precision, and F1 score are defined in the supplementary material.

An image interpretation session was also conducted for the 49 cases by another board-certified radiologist without referring to the MRI images of all the slices. For each patient, (1) the radiologist visually evaluated the head CT images and recorded the positions at which AIS was suspected, and (2) the results of our detection system were disclosed to the radiologist, following which the radiologist could alter his results if required. The radiologist used the results generated by the two-stage model along with the FPR model that involved three-slice images as the detection results of the software system.

Statistical analysis

The sensitivity of the radiologist's results with and without the detection results of the software system was compared, and the McNemar test was used to assess the statistical significance of the difference in sensitivity. P-values of less than 0.05 were considered statistically significant. R (version 3.6.0, available at http://www.R-project.org/) and exact2x2 package (version 1.6.3.1) were used for the statistical analysis.

**Results**

A summary of the detection results of the radiologist and software system are presented in Table 1, and the raw results of the radiologist's image interpretation are presented in Table S1 of the supplementary material. The sensitivity was as follows: one-stage model, 42.7%; two-stage model with the FPR model involving a one-slice image, 30.7%; two-stage model with the FPR model involving three-slice images, 37.3%; radiologist without software system, 33.3%; and radiologist with software system, 41.3%. The FPCs were as follows: one-stage model, 2.837; two-stage model with the FPR model involving a one-slice image, 2.102; two-stage model with the FPR model involving three-slice images, 1.265; radiologist without software system, 0.327; and radiologist with software system, 0.388. As shown in Table 1, although the sensitivity of the one-stage model was the best among the other models, its FPC was the worst. While Table 1 shows that the FPR model was effective in reducing FPs, the effectiveness of the two-stage model with the FPR model involving three-slice images was better than that of the FPR model involving one-slice image. Therefore, we used the two-stage model with the FPR model involving three-slice images for the image interpretation session.

**Table 1.**

| Software/Radiologist | Sensitivity (%) | FPC | Precision (%) | F1 score |
|---|---|---|---|---|
| One-stage model (YOLOv3 only) | 42.7 | 2.837 | 18.7 | 0.260 |
| Two-stage model with FPR model involving one-slice image | 30.7 | 2.102 | 18.2 | 0.229 |
| Two-stage model with FPR model involving three-slice images | 37.3 | 1.265 | 31.1 | 0.339 |
| Board-certified radiologist without software | 33.3 | 0.327 | 61.0 | 0.431 |
| Board-certified radiologist with software | 41.3 | 0.388 | 62.0 | 0.496 |

Summary of detection results by software and a board-certified radiologist

Abbreviations: false positive reduction (FPR), false positive (FP), FPs per one case (FPC), You Only Look Once v3 (YOLOv3).

According to Table S1 in the supplementary material, the radiologist added nine new suspected lesions to the radiologist's detection results following the disclosure of the software results. The suspected lesions that were marked before the disclosure were not deleted following the disclosure. Among the nine new lesions, six and three lesions were TPs and FPs, respectively. Table 2 presents the confusion matrix representing the image interpretation results of the radiologist with and without the detection results of the software system. On adding six TPs, the difference between the radiologist's sensitivity with and without the software system (41.3% v.s. 33.3%) was significant (p-value = 0.0313). The representative CT images and software detection results are depicted in Figures 3 and 4.

**Table 2.**

|  |  | Result of board-certified radiologist with software | |
|---|---|---|---|
|  |  | TP | FN |
| Result of board-certified radiologist without software | TP | 25 | 0 |
|  | FN | 6 | 44 |

Confusion matrix of image interpretation results by a board-certified radiologist with and without detection results of software

Note: The radiologist did not delete his detection results after the disclosure of software results.

Abbreviations: true positive (TP), false negative (FN).

**Figure 3**

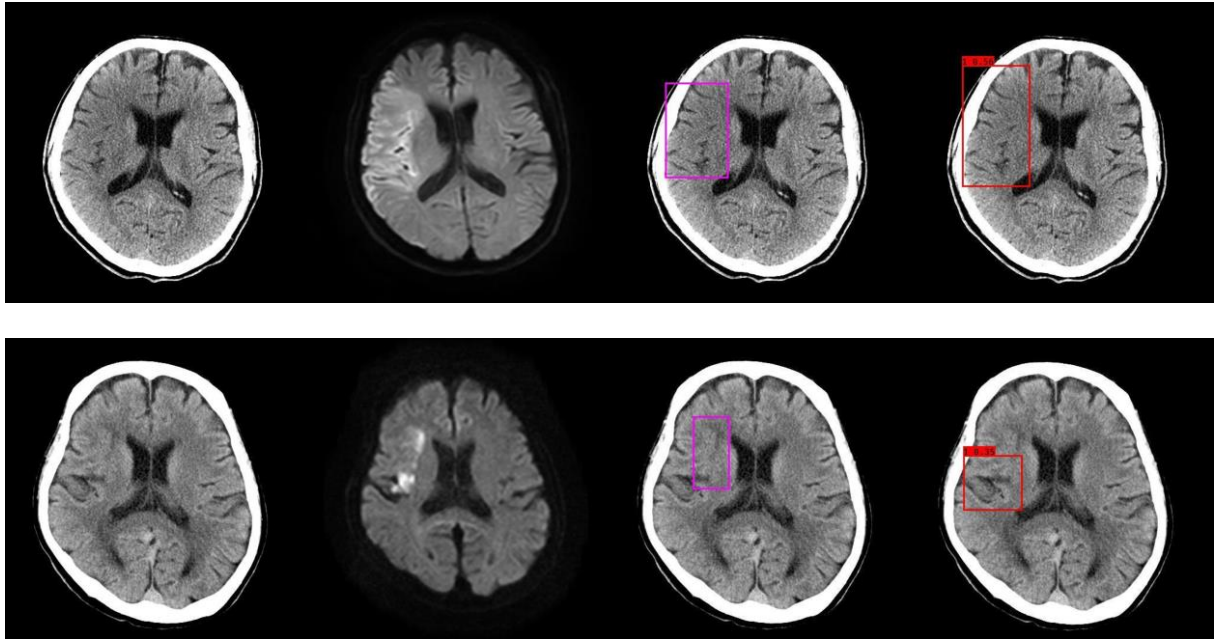

Two cases of acute ischemic stroke that a radiologist could not detect without the aid of the software system but was able to with the software system

Note: CT image, DWI, CT image with ground truth, and CT image with the software detection result are displayed from left to right. The purple and red rectangular boxes represent the ground truth and software detection result indicating acute ischemic stroke.

Abbreviations: diffusion-weighted image (DWI), computed tomography (CT)

**Figure 4**

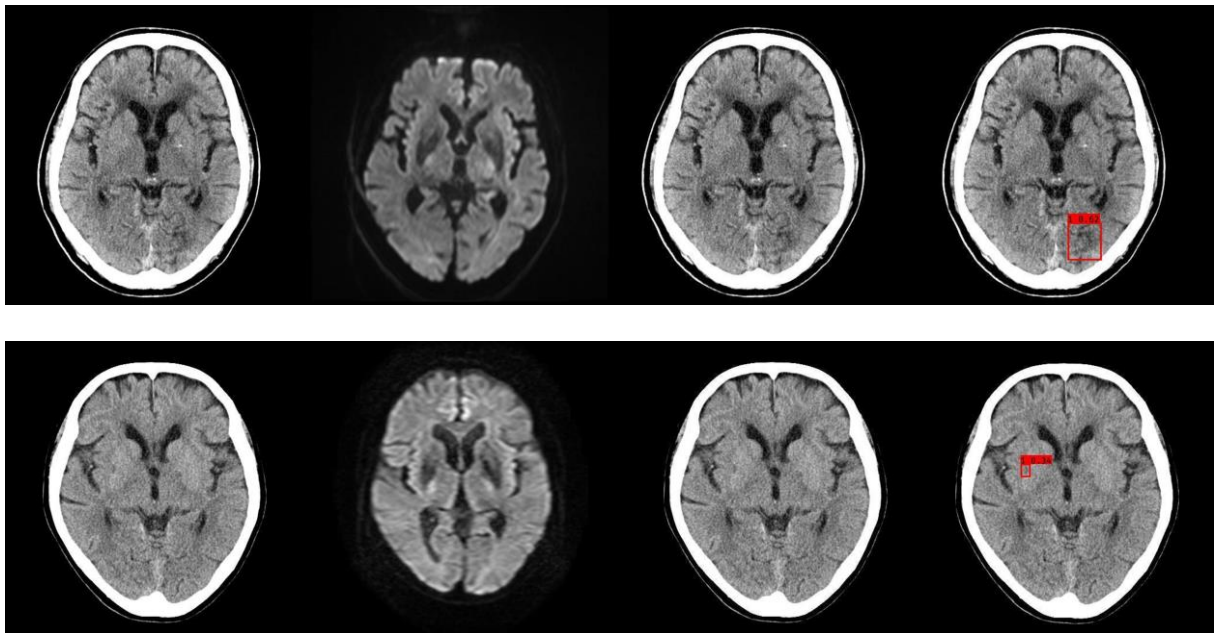

Two cases of false positive radiologist results caused by the software detection result

Note: CT image, DWI, CT image with ground truth (no acute ischemic stroke), and CT image with the software detection result are displayed from left to right. The red rectangular box represents the software detection result.

Abbreviations: diffusion-weighted image (DWI), computed tomography (CT)

## **Discussion**

The results of the current study demonstrate that the two-stage deep learning model was able to detect AIS more accurately than the one-stage model. Although the sensitivity of AIS detection was the highest in the one-stage model involving the use of YOLOv3, the number of FPs was also the highest. Therefore, it was considered inappropriate for medical doctors to use this one-stage model for the interpretation of head CT images. In order to reduce the number of FPs associated with the use of YOLOv3, we constructed a two-stage model of deep learning that was a combination of the YOLOv3 model and VGG16-based FPR model, and the resulting two-stage model efficiently reduced the number of FPs. In addition, the FPR model that required three consecutive slices as its input was more effective in reducing the FP numbers than the FPR model that required one slice. It is important to note that when AIS detection was performed by the radiologist along with the software system, nine new lesions (six TPs and three FPs) were added to the results compared to when the process was performed without the software system. The addition of these new lesions significantly improved the sensitivity of the radiologist in the detection of AIS.

The main purpose of the current study was to develop the AIS detection software on non-contrast CT which were useful for radiologists. The novelty of the current study can be explained via two points: (i) our detection system for examining CT images focused on MRI-detectable AIS, and (ii) our system was able to assist a radiologist in the detection of MRI-detectable AIS. Regarding the first point, our results demonstrated that when MRI was used as the gold standard, the radiologist's sensitivity in the detection of AIS on CT images was low (radiologist's sensitivity = 33.3%). This low sensitivity is compatible with that observed in previous studies. Therefore, a CT dataset that was validated using MRI data was successfully constructed in the current study. Concerning the second point, our detection software significantly improved the sensitivity of the radiologist's AIS detection (from 33.3% to 41.3%).

In the current study, we constructed the MRI-validated AIS CT dataset and used it to develop our AIS detection software. Two previous studies have demonstrated that deep learning models could reliably identify intracranial hemorrhage [11,16]. In addition,

Prevedello *et al.* demonstrated that a deep learning model could accurately detect intracranial hemorrhage and AIS during the initial assessment [9]. The former of the two studies did not focus on AIS, and the latter used a CT dataset that was not validated by MRI. These three studies did not evaluate whether or not their software could assist clinicians. On the other hand, we evaluated whether our AIS detection system on non-contrast CT was useful for radiologists.

As an input for the FPR model, inputting consecutive three-slice CT images was more effective than inputting a one-slice CT image. When we visually evaluated the detection results of YOLOv3, we found that the partial volume effect of the CT images [22] was frequently detected as an FP. The partial volume effect is an artifact that occurs when the slice thickness of a CT image is large; however, it is difficult to avoid this artifact on clinical CT images. To evaluate the partial volume effect in clinical practice, it is necessary to visually evaluate consecutive CT images that include the partial volume effect. In order to incorporate this visual evaluation method into the FPR model, our FPR model based on VGG16 in the current study used consecutive three-slice CT images as its input. According to data presented in Table 1, this method was effective in reducing the number of FPs, which demonstrates its usefulness.

Because this study mainly focused on the development of an AIS detection system, the clinical evaluation of the system was limited. For example, the radiologist's FPs that were caused by the system were not fully assessed. In the current study, because the number of TPs generated by the system was higher than that of FPs, we expect that the benefits of this system would exceed any disadvantages resulting from its use. However, this must be clinically investigated.

The results presented in Table 1 demonstrate that the radiologist's FPs that were caused by the software system were minimal and that most of the FPs produced by our system were correctly classified by the radiologist. These results show that the interaction between the software system and radiologist might lead to improving the detectability of AIS. The interaction between our software system and medical doctors should be investigated in future research.

There are several limitations of the present study. First, the number of radiologists who participated in the current study was small. The efficacy of our detection system should be evaluated by a larger number of radiologists. Second, external validation was not performed. Because our dataset was obtained from two different institutions, our detection system appears to be robust when faced with variations in datasets. However, this must be confirmed using an external validation set. Last, while our results show that the lesion-based sensitivity was significantly improved, patient-based sensitivity was not (supplementary material). Dzialowski et al. and Hirano et al. show that extent of early ischemic changes caused by AIS was related with intracranial hemorrhage in patients treated with tissue plasminogen activator [23,24]. We expect that the higher lesion-based sensitivity can be useful for avoiding intracranial hemorrhage in patients treated with tissue plasminogen activator. For this sake, our software might be useful. The three limitations should be addressed in future studies.

## **Conclusions**

To the best of our knowledge, there was no study to evaluate whether deep-learning-based AIS detection system on non-contrast computed tomography was useful for a board-certified radiologist. The results of the current study demonstrate that our detection system involving a two-stage deep learning model could significantly improve the sensitivity of radiologist in the detection of AIS.

**Sources of funding**

The present study was supported by the Japan Society for the Promotion of Science (JSPS) KAKENHI (Grant Number JP19K17232 and 17J07699). The funder had no role in the present study.

**Conflict of interest statement**

None

**References**

1. The top 10 causes of death. [cited 8 Jan 2020]. Available: https://www.who.int/news-room/fact-sheets/detail/the-top-10-causes-of-death

2. Mokin M, Ansari SA, McTaggart RA, Bulsara KR, Goyal M, Chen M, et al. Indications for thrombectomy in acute ischemic stroke from emergent large vessel occlusion (ELVO): Report of the SNIS Standards and Guidelines Committee. Journal of NeuroInterventional Surgery. BMJ Publishing Group; 2019. pp. 215–220. doi:10.1136/neurintsurg-2018-014640

3. Bluhmki E, Chamorro Á, Dávalos A, Machnig T, Sauce C, Wahlgren N, et al. Stroke treatment with alteplase given 3·0-4·5 h after onset of acute ischaemic stroke (ECASS III): additional outcomes and subgroup analysis of a randomised controlled trial. Lancet Neurol. 2009;8: 1095–1102. doi:10.1016/S1474-4422(09)70264-9

4. Vymazal J, Rulseh AM, Keller J, Janouskova L. Comparison of CT and MR imaging in ischemic stroke. Insights into Imaging. 2012. pp. 619–627. doi:10.1007/s13244-012-0185-9

5. Chalela JA, Kidwell CS, Nentwich LM, Luby M, Butman JA, Demchuk AM, et al. Magnetic resonance imaging and computed tomography in emergency assessment of patients with suspected acute stroke: a prospective comparison. Lancet. 2007;369: 293–298. doi:10.1016/S0140-6736(07)60151-2

6. Powers WJ, Derdeyn CP, Biller J, Coffey CS, Hoh BL, Jauch EC, et al. 2015 American Heart Association/American stroke association focused update of the 2013 guidelines for the early management of patients with acute ischemic stroke regarding endovascular treatment: A guideline for healthcare professionals from the American Heart Association/American stroke association. Stroke. Lippincott Williams and Wilkins; 2015. pp. 3020–3035. doi:10.1161/STR.0000000000000074

7. Akasaka T, Yakami M, Nishio M, Onoue K, Aoyama G, Nakagomi K, et al. Detection of suspected brain infarctions on CT can be significantly improved with temporal subtraction images. Eur Radiol. 2018; 1–11. doi:10.1007/s00330-018-5655-0

8. Gulshan V, Peng L, Coram M, Stumpe MC, Wu D, Narayanaswamy A, et al. Development and validation of a deep learning algorithm for detection of diabetic retinopathy in retinal fundus photographs. JAMA - J Am Med Assoc. 2016;316: 2402–2410. doi:10.1001/jama.2016.17216

9. Prevedello LM, Erdal BS, Ryu JL, Little KJ, Demirer M, Qian S, et al. Automated critical test findings identification and online notification system using artificial intelligence in imaging. Radiology. 2017;285: 923–931. doi:10.1148/radiol.2017162664

10. Cho J, Park KS, Karki M, Lee E, Ko S, Kim JK, et al. Improving Sensitivity on Identification and Delineation of Intracranial Hemorrhage Lesion Using Cascaded Deep Learning Models. J Digit Imaging. 2019;32: 450–461. doi:10.1007/s10278-018-00172-1

11. Kuo W, Häne C, Mukherjee P, Malik J, Yuh EL. Expert-level detection of acute intracranial hemorrhage on head computed tomography using deep learning. Proc Natl Acad Sci U S A. 2019;116: 22737–22745. doi:10.1073/pnas.1908021116

12. Rajpurkar P, Irvin J, Zhu K, Yang B, Mehta H, Duan T, et al. CheXNet: Radiologist-Level Pneumonia Detection on Chest X-Rays with Deep Learning. 2017 [cited 8 Jan 2020]. Available: http://arxiv.org/abs/1711.05225

13. Esteva A, Kuprel B, Novoa RA, Ko J, Swetter SM, Blau HM, et al. Dermatologist-level classification of skin cancer with deep neural networks. Nature. 2017;542: 115–118. doi:10.1038/nature21056

14. Lee H, Lee E-J, Ham S, Lee H-B, Lee JS, Kwon SU, et al. Machine Learning Approach to Identify Stroke Within 4.5 Hours. Stroke. 2020; STROKEAHA119027611. doi:10.1161/STROKEAHA.119.027611

15. Sheth SA, Lopez-Rivera V, Barman A, Grotta JC, Yoo AJ, Lee S, et al. Machine Learning-Enabled Automated Determination of Acute Ischemic Core From Computed Tomography Angiography. Stroke. 2019;50: 3093–3100. doi:10.1161/STROKEAHA.119.026189

16. Chilamkurthy S, Ghosh R, Tanamala S, Biviji M, Campeau NG, Venugopal VK, et al. Deep learning algorithms for detection of critical findings in head CT scans: a retrospective study. Lancet. 2018;392: 2388–2396. doi:10.1016/S0140-6736(18)31645-3

17. Nishio M, Sugiyama O, Yakami M, Ueno S, Kubo T, Kuroda T, et al. Computer-aided diagnosis of lung nodule classification between benign nodule, primary lung cancer, and metastatic lung cancer at different image size using deep convolutional neural network with transfer learning. PLoS One. 2018;13: 1–12. doi:10.1371/journal.pone.0200721

18. Yamashita R, Nishio M, Do RKG, Togashi K. Convolutional neural networks: an overview and application in radiology. Insights Imaging. 2018;9: 611–629. doi:10.1007/s13244-018-0639-9

19. Wang K, Shou Q, Ma SJ, Liebeskind D, Qiao XJ, Saver J, et al. Deep Learning Detection of Penumbral Tissue on Arterial Spin Labeling in Stroke. Stroke. 2019. doi:10.1161/strokeaha.119.027457

20. Redmon J, Farhadi A. YOLOv3: An Incremental Improvement. 2018 [cited 10 Jan 2020]. Available: http://arxiv.org/abs/1804.02767

21. Simonyan K, Zisserman A. Very deep convolutional networks for large-scale image recognition. 3rd International Conference on Learning Representations, ICLR 2015 - Conference Track Proceedings. International Conference on Learning Representations, ICLR; 2015.

22. Goodenough D, Weaver K, Davis D, LaFalce S. Volume averaging limitations of computed tomography. Am J Roentgenol. 1982;138: 313–316. doi:10.2214/ajr.138.2.313

23. Dzialowski I, Hill MD, Coutts SB, Demchuk AM, Kent DM, Wunderlich O, et al. Extent of early ischemic changes on computed tomography (CT) before thrombolysis: Prognostic value of the Alberta Stroke Program early CT score in ECASS II. Stroke. 2006;37: 973–978. doi:10.1161/01.STR.0000206215.62441.56

24. Hirano T, Sasaki M, Tomura N, Ito Y, Kobayashi S. Low Alberta stroke program early computed tomography score within 3 hours of onset predicts subsequent symptomatic intracranial hemorrhage in patients treated with 0.6 mg/kg alteplase. J Stroke Cerebrovasc Dis. 2012;21: 898–902. doi:10.1016/j.jstrokecerebrovasdis.2011.05.018

**Supplementary material 1: Definition of IoU**

Ground truth (GT) is defined by a radiologist as bounding box of AIS on head CT image. IoU is defined as ratio of intersection and union of predicted bounding box and GT bounding box as follows:

$$IoU = \frac{area|B_P \cap B_{GT}|}{area|B_P \cup B_{GT}|},$$

where $B_P$ is the predicted bounding box and $B_{GT}$ is the GT bounding box.

**Supplementary material 2: Definition of F1 score**

F1 score is defined as the following equations:

$$Precision = \frac{TP}{TP + FP},$$

$$Recall = \frac{TP}{TP + FN},$$

$$F1\ score = \frac{2\ Recall \cdot Precision}{Recall + Precision},$$

where TP, FP, and FN are true positive, false positive, and false negative, respectively. Recall is identical to sensitivity. The probability of AIS was binarized based on threshold, and TP, FP, and FN were calculated for F1 score of our detection system.

**Supplementary material 3: Detail of YOLOv3 model**

YOLOv3 is one-stage detection model of deep learning, which directly predicts the class probability and spatial coordinates of the detection targets. In the current study, YOLOv3 model predicted the probability and spatial coordinates of AIS. YOLOv3 model extracts image feature from its input image, and then predicts the probability and spatial coordinates based on the extracted image feature. To train YOLOv3 model, regression between prediction of YOLOv3 model and ground truth is performed for the probability and spatial coordinates. YOLOv3 uses Darknet-53 for image feature extraction, which has 53 convolutional/fully-connected layers trained on ImageNet. For detection, more layers are stacked onto Darknet-53 to construct network architecture of YOLOv3. YOLOv3 predict the probability and spatial coordinates of detection targets at three different scales. For example, if input size is 416×416, output sizes of YOLOv3 at three different scales are 52×52, 26×26, and 13×13. The output of YOLOv3 at each scale is defined as collection of cells, each cell has bounding boxes, and each bounding box has (i) its spatial coordinates, (ii) object confidence, and (iii) probabilities of detection target classes. Final prediction results of YOLOv3 were obtained by merging the output of YOLOv3 at three different scales.

**Supplementary material 4: Table S1. Raw results of image interpretation by a board-certified radiologist**

| patient case index | N of AIS | N of suspected lesions without software | N of true positives without software | N of false positives without software | N of false negatives without software | N of added suspected lesions with software |
|---|---|---|---|---|---|---|
| 0 | 1 | 0 | 0 | 0 | 1 | 0 |
| 1 | 1 | 0 | 0 | 0 | 1 | 0 |
| 2 | 2 | 0 | 0 | 0 | 2 | 0 |
| 3 | 1 | 0 | 0 | 0 | 1 | 0 |
| 4 | 3 | 1 | 1 | 0 | 2 | 0 |
| 5 | 1 | 1 | 1 | 0 | 0 | 0 |
| 6 | 2 | 1 | 1 | 0 | 1 | 0 |
| 7 | 3 | 2 | 2 | 0 | 1 | 0 |
| 8 | 1 | 0 | 0 | 0 | 1 | 0 |
| 9 | 1 | 2 | 1 | 1 | 0 | 1 (TP) |
| 10 | 1 | 0 | 0 | 0 | 1 | 0 |
| 11 | 3 | 0 | 0 | 0 | 3 | 0 |
| 12 | 1 | 4 | 1 | 3 | 0 | 0 |
| 13 | 3 | 3 | 2 | 1 | 1 | 1 (TP) |
| 14 | 1 | 1 | 0 | 1 | 1 | 1 (FP) |
| 15 | 0 | 0 | 0 | 0 | 0 | 0 |
| 16 | 2 | 0 | 0 | 0 | 2 | 0 |
| 17 | 3 | 3 | 3 | 0 | 0 | 0 |
| 18 | 1 | 3 | 1 | 2 | 0 | 0 |
| 19 | 1 | 0 | 0 | 0 | 1 | 0 |
| 20 | 5 | 3 | 2 | 1 | 3 | 0 |
| 21 | 1 | 1 | 1 | 0 | 0 | 0 |
| 22 | 1 | 1 | 1 | 0 | 0 | 1 (TP) |
| 23 | 1 | 1 | 1 | 0 | 0 | 1 (TP) |
| 24 | 0 | 0 | 0 | 0 | 0 | 0 |
| 25 | 2 | 1 | 1 | 0 | 1 | 0 |
| 26 | 1 | 1 | 1 | 0 | 0 | 0 |
| 27 | 0 | 0 | 0 | 0 | 0 | 0 |
| 28 | 1 | 1 | 0 | 1 | 1 | 0 |
| 29 | 0 | 1 | 0 | 1 | 0 | 1 (FP) |
| 30 | 3 | 0 | 0 | 0 | 3 | 0 |
| 31 | 2 | 1 | 1 | 0 | 1 | 0 |
| 32 | 2 | 1 | 1 | 0 | 1 | 0 |
| 33 | 1 | 0 | 0 | 0 | 1 | 0 |
| 34 | 0 | 0 | 0 | 0 | 0 | 0 |
| 35 | 2 | 1 | 1 | 0 | 1 | 0 |
| 36 | 0 | 0 | 0 | 0 | 0 | 0 |
| 37 | 1 | 0 | 0 | 0 | 1 | 0 |
| 38 | 3 | 1 | 1 | 0 | 2 | 1 (TP) |
| 39 | 4 | 3 | 1 | 2 | 3 | 0 |
| 40 | 1 | 1 | 1 | 0 | 0 | 0 |
| 41 | 1 | 3 | 1 | 2 | 0 | 0 |

| 42 | 2 | 1 | 1 | 0 | 1 | 0 |
| 43 | 3 | 2 | 1 | 1 | 2 | 1 (TP) |
| 44 | 2 | 2 | 0 | 2 | 2 | 1 (FP) |
| 45 | 1 | 1 | 1 | 0 | 0 | 0 |
| 46 | 2 | 0 | 0 | 0 | 2 | 0 |
| 47 | 1 | 2 | 1 | 1 | 0 | 0 |
| 48 | 0 | 0 | 0 | 0 | 0 | 0 |

Abbreviations: true positive (TP), false positive (FP), acute ischemic stroke (AIS).